\documentclass[12pt]{article}
\usepackage{pslatex} 
\usepackage{amsmath,amssymb,amsfonts}
\usepackage{tabularx}
\usepackage{graphicx}
\usepackage[margin=2.5cm]{geometry}
\usepackage{epstopdf}
\usepackage{setspace}
\usepackage{authblk}
\usepackage{multirow}
\usepackage[pdfpagemode={UseOutlines},bookmarks=true,bookmarksopen=true,
   bookmarksopenlevel=0,bookmarksnumbered=true,hypertexnames=false,
   colorlinks,linkcolor={blue},citecolor={green},urlcolor={blue},
   pdfstartview={FitV},unicode,breaklinks=true]{hyperref}
\usepackage{breqn}
\usepackage{pdflscape}
\usepackage[table]{xcolor}
\usepackage{color}
\usepackage{tikz}
\usetikzlibrary{shapes,arrows}
\usepackage{verbatim}
\newcommand{\bj}[1]{\textcolor{blue}{#1}}

\begin{document}

\title{Exploration of mechanical, thermal conduction and electromechanical properties of graphene nanoribbon springs}
\author{Brahmanandam Javvaji$^{a}$, Bohayra Mortazavi $^{a}$, Timon Rabczuk$^{b}$ and Xiaoying Zhuang$^{a,*}$ \\
		\emph{\small{$^a$Institute of Continuum Mechanics, Leibniz Universit\"at Hannover, Appelstr. 11, 30167 Hannover, Germany}} \\
		\emph{\small{$^b$Institute of Structural Mechanics, Bauhaus University Weimar, Marienstrasse 15, 99423 Weimar, Germany.}} \\
		\small{$^*$Corresponding author E-mail:~zhuang@ikm.uni-hannover.de}
}
\date{}
\maketitle
\doublespacing
\section*{Abstract}
Recent experimental advances [Liu \textit{et al., npj 2D Materials and Applications}, 2019, \textbf{3}, 23] propose the design of graphene nanoribbon spring (GNRS) to substantially enhance the stretchability of pristine graphene. GNRS is a periodic undulating graphene nanoribbon, where undulations are of sinus or half-circles or horseshoe shapes. Besides those, GNRS geometry depends on design parameters, like pitch's length and amplitude, thickness and joining angle. Because of the fact that parametric influence on the resulting physical properties are expensive and complicated to be examined experimentally, we explore the mechanical, thermal and electromechanical properties of GNRS  using molecular dynamics simulations. Our results demonstrate that horseshoe shape design of GNRS (GNRH) can distinctly outperform the graphene kirigami design concerning the stretchability. The thermal conductivity of GNRS were also examined by developing a multiscale modeling, which suggests that the thermal transport along these nanostructures can be effectively tuned. We found that however, the tensile stretching of GNRS and GNRH does not yield any piezoelectric polarization. The bending induced hybridization change results in a flexoelectric polarization, where the corresponding flexoelectric coefficient is $25\%$ higher than graphene. Our results provide a comprehensive vision to the critical physical properties of GNRS and may help to employ the outstanding physics of the graphene to design novel stretchable nanodevices.

\section{Introduction}
During the last decade, graphene \cite{Novoselov2004,Geim2007,CastroNeto2009} the two-dimensional (2D) form of sp$^2$ carbon atoms has attracted astonishing attentions of scientific and industrial communities, stemmed from its extraordinarily high mechanical \cite{Lee2008} and thermal \cite{Balandin2011} conduction properties along with exceptional electronic and optical characteristics. In particular, graphene can exhibit remarkably high Young's modulus and tensile strength of 1000 GPa and 130 GPa \cite{Lee2008} respectively, along with superior thermal conductivity of around 4000 W/mK \cite{Ghosh2010} that outperforms diamond and other conventional materials. The exceptional physical and chemical properties of graphene, promoted the experimental and theoretical endeavours to fabricate novel graphene's 2D counterparts, and subsequently explore their intrinsic properties and application prospects. As a results of scientific accomplishments, 2D materials family is commonly considered as the most vibrant class of materials, in which new members are continuously introduced, either theoretically predicted or experimentally fabricated. Worthy to remind that despite the outstanding properties of pristine graphene, it is not an ideal material and naturally shows few drawbacks, such as the lack of an electronic band-gap and brittle failure mechanism \cite{Lee2008,Zhang2014}. In addition, the ultra-high thermal conductivity of graphene also prohibits its application for thermoelectric energy generation. \\
Nonetheless, it has been also practically proven that graphene can show largely/finely tunable electronic, mechanical, thermal, optical and electromechanical properties, with accurately controlled mechanical straining \cite{Guinea2012,Metzger2010,Pereira2009,Barraza-Lopez2013,Guinea2010,javvaji2016mechanical}, defect engineering \cite{Lherbier2012,Cummings2014,Cresti2008,Javvaji2018} or chemical doping \cite{Wehling2008,Miao2012,Wang2008,Schedin2007,Soriano2015}. We remind that for many centuries, springs have been playing pivotal roles in the design and fabrication of various kind of devices. The importance of springs originates from the fact that, while the mechanical properties of a material is considered as its inherent property and thus invariable, when it is shaped in the form of springs the subsequent structures can show superior stretchability and diverse mechanical responses. In particular, the design of spring like structures have been known as one of the most effective ways to design highly stretchable and flexible moving components.\\
For the employment of graphene in flexible nanoelectronics, its ductile and highly rigid mechanical properties are undesirable. Therefore, engineering of the graphene design in order to improve its stretchability is a critical issue \cite{Feng2018,Hu2019,Liu2019c,Jia2019,Yang2019}. To address this challenge, in a most recent exciting experimental advance by Liu et al.~\cite{Liu2019} the old idea of spring design has been applied for the case of graphene to enhance its stretchability and flexibility via a nanowire lithography technology. This experimental advance consequently raises questions concerning that how the design of graphene springs can be improved to reach higher degrees of stretchability. In addition, the electronic and heat transport properties of these novel nanostructures should be also examined, in order to provide comprehensive visions for the design of nanodevices. As a common challenge in the electronic apparatus, the thermal conductivity of employed components should be high to effectively dissipate the excessive heats. On the opposite side, low thermal conductivity is a key requirement for the enhancement in the efficiency of thermoelectric energy conversion. As a common barrier, it is well known that the evaluation of the mechanical and transport properties of graphene springs by the experimental tests are not only complicated but also time consuming and expensive as well. This study therefore aims to investigate the mechanical response and heat conduction properties of graphene springs via conducting extensive classical molecular dynamics simulations. Since the graphene is the frontier and symbolic member of 2D materials, commonly the experimental and theoretical methodologies that are applied for the graphene can be extended for the other members of 2D materials family. We are thus hopeful that the obtained results by this study may sever as valuable guides for the future theoretical and experimental studies on the design of 2D materials spring like structures.
\section{Methods}
We conducted classical molecular dynamics (MD) simulations to evaluate the mechanical properties and thermal conductivity of graphene springs, using the large scale atomic/molecular massively parallel simulatior package \cite{Plimpton1995}. To this aim, we used the optimized Tersoff potential parameter set proposed by Lindsay and Broido \cite{Lindsay2010} for introducing the atomic interaction between carbon. This version of Tersoff potential is not only highly computationally efficient, but also can yield accurate results for the mechanical and thermal properties of graphene. We analysed the mechanical response by conducting the uniaxial tensile simulations at room temperatures. For the evaluation of mechanical properties, we modified the cutoff of Tersoff potential from 0.18 nm to 0.20 nm to remove an unphysical stress pattern and moreover accurately reproduce the experimental results for the tensile strength of pristine graphene \cite{Mortazavi2016}. In this case, the time increment of MD simulations was set at 0.25 fs. Before applying the loading conditions, all structures were equilibrated using Nos\'{e}-Hoover thermostat method. For the loading condition, a constant engineering strain rate of 1$\times$10$^8$ s$^{-1}$ was applied, by increasing the periodic size of the simulation box along the loading direction in every time step. Virial stresses at every step were recorded and averaged over every 20 ps intervals to report the stress-strain relations. 
To evaluate the thermal conductivity, we used equilibrium molecular dynamics (EMD) method. The time increment of EMD simulations was set at 0.25 fs. In the EMD method, the heat flux vector was calculated via:
\begin{equation}
\mathbf{J}(t) = \sum_i \left( e_i \mathbf{v}_i + \frac{1}{2} \sum_{i < j} \left(\mathbf{f}_{ij} \cdot \left(\mathbf{v}_i + \mathbf{v}_j\right)\right)\mathbf{r}_{ij}\right)
\end{equation}	
where $e_i$ and $\mathbf{v}_i$ are respectively the total energy and velocities of atom $i$, $\mathbf{f}_{ij}$ and $r_{ij}$ are the interatomic force and position vector between atoms $i$ and $j$, respectively. In the approach, first the structures were equilibrated at a constant volume and room temperature using the Berendsen thermostat method. Before the evaluation of thermal conductivity, in order to remove the effects of previously applied thermostat, we used constant energy (NVE) simulations. For the evaluation of effective thermal conductivity, individual simulations were conducted for 1 ns under the NVE ensemble. The EMD method relies on relating the ensemble average of the heat current auto-correlation function (HCACF) to the thermal conductivity $k$, via the Green-Kubo expression:
\begin{equation}
k_{\alpha \beta} = \frac{1}{Vk_BT^2} \int_0^{\infty} \left\langle J_{\alpha} (0) J_{\beta} (t) \right \rangle dt
\end{equation}
where $k_B$ is Boltzmann's constant, $T$ is the simulation temperature, and $V$ is the total volume of the graphene spring. In order to reach a converged thermal conductivity, several independent simulations with uncorrelated initial velocities were carried out and the acquired HFACFs were averaged. \\
\section{Results}	
\begin{figure}[h]
	\centering
	\includegraphics[width=1.0\linewidth]{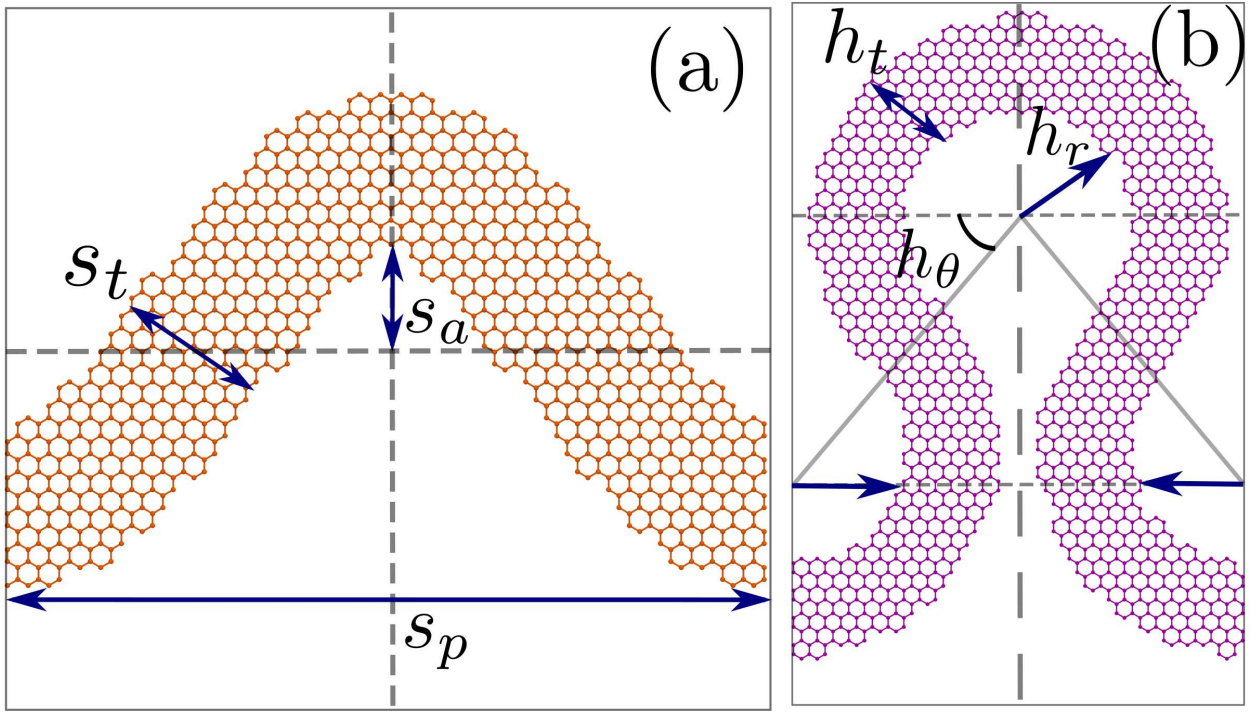}
	\caption{Unit cell representation and definition of various structural parameters for (a) sinus shape and (b) horseshoe shape graphene nanosprings.}
	\label{fig:ucell}
\end{figure}
Fig.~\ref{fig:ucell}(a) and (b) shows the atomic unit-cell representation for graphene nanoribbon (GNR) in sinus shape and horseshoe shape, respectively. Sinus shape of GNR (GNRS) obtained from cutting the infinite graphene sheet using two sine curves that are parallel to each other. The parameters defining the sine curve are pitch length $(s_p)$ and amplitude $(s_a)$. The locus of each point normals with a constant length $(s_t)$ creates a parallel sine curve. The variable $s_t$ defines the thickness of GNRS. The choice of $(s_p)$ and $(s_a)$ is arbitrary. However, $(s_t)$ should be less than the radius of the curvature of the sine curve, which is to avoid producing bigger arcs that cover the crests and troughs (cusps) of the sine curve. Using these variables for GNRS, we define a sinus shaped region on a pristine graphene sheet and removed atoms other than this region, which creates an initial atomic configuration for GNRS. The volume of GNRS is defined as the area under the parallel sine curve times the thickness of pristine graphene sheet $3.3~\text{\AA}$ \cite{Ishigami2007}. \\
The horseshoe shape design of GNR (GNRH) composes by connecting two circular arcs of the inner radius $(h_r)$ with the intersecting angle $(h_{\theta})$. When $h_{\theta} = 0^\circ$, GNRH looks like a series of connected semi-circles and for $h_{\theta} > 0^\circ$ and $h_{\theta} \leq 45^\circ$, the horseshoe design is obtained.  For $h_{\theta}$ more than $45^\circ$, the semi-circles merge each other, which is not desirable. These parameters define the shape of a single horseshoe curve. Another curve with radius $h_r + h_t$ creates a parallel curve, where $h_t$ defines the thickness of GNRH. We construct the horseshoe-shaped region on a pristine graphene sheet and removed atoms other than this region, which creates an initial atomic configuration for the GNRH system. \bj{After the initial preparation of spring structures, we removed the carbon atoms bonded with a single carbon atom along the lateral edges. Since these atoms can lead to instability in simulations.} It is worth noting that the carbon atoms belong to the curved edges of the GNRS and GNRH system are not terminated with hydrogen atoms. \\
With the defined geometrical parameters and initial atomic configurations for GNRS and GNRH, we consider the following parameter sets for estimating the mechanical and thermal properties. \textbf{Parameter set $\mathbf{s_p-s_t-s_a}$:} The starting value for $s_p$ is 9 nm (where the size effects are absent in pristine graphene \cite{javvaji2016mechanical}), and $s_a$ is 2.5 nm, where at least 10 carbon rings accompanying in the lateral direction. The minimum possible value of $s_t$ is 1.6 nm for this combination of $s_p$ and $s_a$ for having a reasonable thickness for these spring systems. Further, we increase the values of $sp$, $st$ and $sa$ by integral multiples from 1 to 5. \textbf{Parameter $\mathbf{h}_{\theta}$:} We consider $h_\theta$ as $0^\circ, 15^\circ, 30^\circ$ and $45^\circ$ by keeping $h_r$ at 2.5 nm and $h_t$ as 1.6 nm. Using this choice of parameters, we explore the effect of $h_\theta$ on mechanical and thermal properties. \textbf{Parameter $\mathbf{h_r-h_t}$:}  In this set, $h_r$ and $h_t$ parameters are scaled by integral multiples from 1 to 5 starting from 2.5 and 1.6 nm at fixed $h_\theta$. We fix the value of $h_\theta$ from previous parameter set, which has shown exceptional mechanical properties. \bj{The spring structures from the different parameter sets are made by cutting from graphene sheets oriented along the zigzag direction. Our comparative results discussed in the supplementary information\dag document however confirm that the orientation of original graphene sheet show negligible effects in the estimated properties.} 
\subsection{Mechanical properties}
\begin{figure*}[h]
	\centering
	\includegraphics[width=1.0\linewidth]{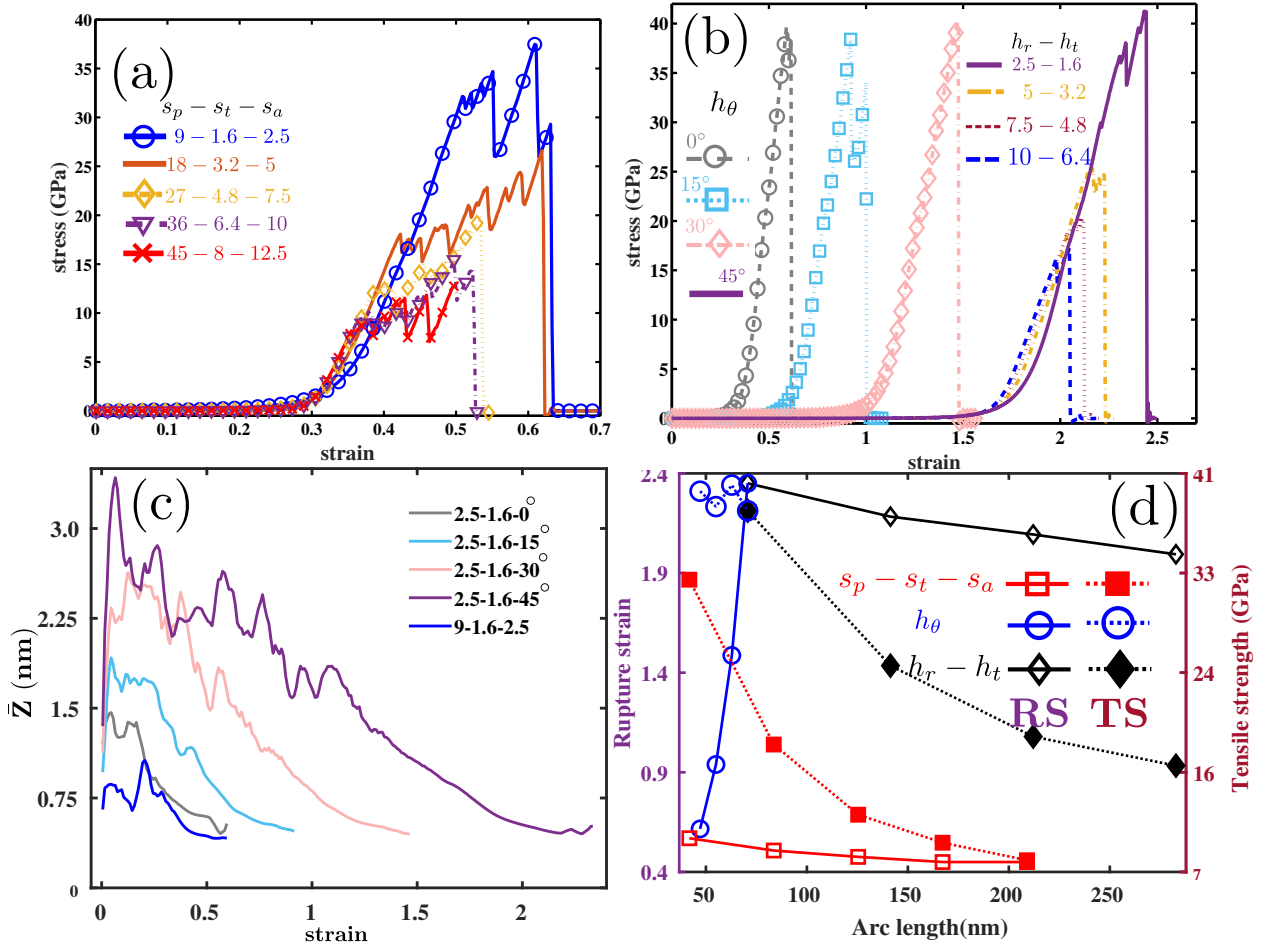}
	\caption{Stress-strain curves for (a) parameter set $s_p-s_t-s_a$ in GNRS, (b) $h_{\theta}$ and $h_r-h_t$ for GNRH. (c) Standard deviation of the $z-$coordinates for selected GNRS and GNRH systems. (d) Variation of rupture strain and tensile strength (TS) for GNRS and GNRH systems with arc length. Solid lines indicate RS and dashed lines correspond to TS.}
	\label{fig:SS}
\end{figure*}
\begin{figure*}
	\centering
	\includegraphics[width=1.0\linewidth]{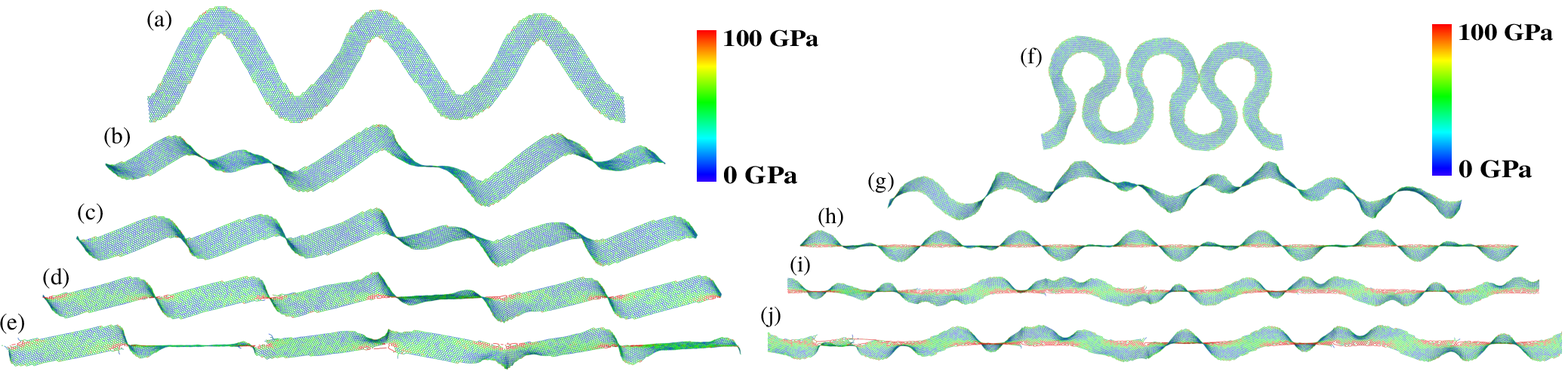}
	\caption{Tensile deformed atomic configurations for (a-e)  $18-3.2-5$ GNRS and (f-j) $5-3.2-45^\circ$ GNRH. The strain levels for the atomic snapshots are as follows: (a and f) 0.0, (b) 0.18, (c) 0.30, (d) 0.42, (e) 0.54, (g) 1.42, (h) 2.0, (i) 2.17 and (j) 2.23. Color bar indicate the stress values.}
	\label{fig:ss-ac}
\end{figure*}
Fig.~\ref{fig:SS}(a) shows the stress-strain curves for the parameter set $s_p-s_t-s_a$. Under stretching of GNRS system, the stress values remain low up to a strain of 0.3. This behavior is in contrast to that of pristine graphene. The given deformation stretches the bond lengths in pristine graphene and increases the stress levels. In GNRS systems, the given deformation deflects the atomic system instead of stretching the bonds, which is similar to the earlier observations for graphene kirigami \cite{Mortazavi2017}. \bj{Further, the recent study on nonlinear vibrations for helical graphene nanoribbon shows a transformation of softening type to hardening type response between the amplitude to frequency variation during the increase in mechanical loading\cite{Liu2020}. In GNRS, the initial plateau in stress-strain curve up to a strain of $0.3$ corresponds to the soft nonlinearity. Because of the transformation from softening to hardening type natural vibrations, GNRS system start stretching due to the increased loading.} Fig.~\ref{fig:ss-ac}(a) visualizes the initial (strain is 0) atomic configuration for $18-3.2-5$ GNRS. Visual molecular dynamics package \cite{Humphrey1996} has been used to generate the atomic snapshots. At a strain of 0.18, because of the deflections, a higher number of crests and troughs are observed in Fig.~\ref{fig:ss-ac}(b).  Further straining to 0.30, deflects the GNRS without increasing the peaks (Fig.~\ref{fig:ss-ac}(c)). For strain less than 0.30, there is no significant increase in the stress distributions. When strain is more than 0.30, there is an increase in stress due to the bond stretching. Fig.~\ref{fig:ss-ac}(d) shows the locations of stress concentrations (near peaks) in the atomic configurations. Atomic configuration in Fig.~\ref{fig:ss-ac}(e) shows the multiple bond failures at strain 0.54. The failure process for other members in this parameter set is similar to the configuration $18-3.2-5$ GNRS. However, there are differences in the tensile strength (TS) and onsite of failure/rupture strain (RS) values due to the changes in the geometrical parameters. We use arc length as the variable to discuss the parametric dependence of RS and TS on $s_p$, $s_t$ and $s_a$. Fig.~\ref{fig:SS}(d) shows that RS is nearly converged to 0.45 for arc lengths larger than 170 nm. Whereas, TS has a very large variation with arc length, from 33 to 7 GPa, which is due to the increase of parameter $s_t$ in GNRS design. As $s_t$ increases, the interaction between stress centers near crest and troughs of GNRS decreases, which decrease the total system stress. Further increase in the size parameters will converge to a constant value. Finally, for GNRS systems, the stress concentrates near the peak portions lead to global failure by breaking the bonds in carbon rings. The RS for the experimentally manufactured GNRS is about $0.35$ for a system with thickness $50$ nm \cite{Liu2019}. This observation is in close agreement with the converged RS value of $0.43$ for spring design, which shows that our simulation predictions are accurate enough to explore the theoretical understanding for these novel design. \\
Fig.~\ref{fig:SS}(b) shows the stress-strain response for parameter set $h_\theta$, where $h_r$ and $h_t$ values are at 2.5 and 1.6 nm, respectively. When $h_\theta = 0^\circ$, there is no stress rise up to a strain of 0.3. For strain range 0.3 to 0.6, the deformation in atomic configuration rises the system stress followed by a failure.  The strain range for the non-zero portion of the stress-strain curve shifts between 0.5 to 1.0 when $h_\theta$ is $15^\circ$. For $h_\theta$ equal to $30^\circ$, the non-zero portion of the stress-strain curve span the strain range of 0.94 to 1.5. For 45$^\circ$ GNRH, this strain range increased to 2.4. Figs.~\ref{fig:ss-ac}(f)-(j) shows the atomic configurations for a GNRH with a 45$^\circ$ connecting angle. The closeness between semi-circular segments in GNRH develops strong repulsion interactions compared to GNRS. Such repulsion largely deflects the atomic system. Further, mechanical stretching reduces atomic deflections by maintaining the stress levels via transforming the smooth circular GNRH segments to sharp peaks. Fig.~\ref{fig:ss-ac}(g) shows the atomic configuration with several peaks in GNRH. For strain greater than 1.5, Fig.~\ref{fig:SS}(b) shows a linear stress-strain response due to the bond stretching. At strain 2, GNRH system looks like a combination of thread and knots, as seen in Fig.~\ref{fig:SS}(h), where stress concentrates near the thread portions. Further increase in strain in GNRH, leads to bond failure.  Figs.~\ref{fig:SS}(i) and (j) at strain levels of 2.17 and 2.23 shows the complete failure of GNRH. \\
Interestingly, GNRH with various $h_\theta$ maintained the stress levels when increasing the strain range.
\bj{We consider varying the width parameter $h_t$ in GNRH by keeping the other parameters constant to check its influence on the mechanical properties. The range of $h_t$ is limited by the choice of other two parameters. For example, consider $h_r$ equal to $5$ nm and $h_\theta$ is $45^\circ$. The maximum available value of $h_t$ is $4$ nm. When $h_t$ is greater than $4$ nm, the two circular cross-sections of GNRH unitcell overlap with each other, which is not desirable. We consider $h_t$ values as $1.6$, $2.4$ and $3.2$ and $4$ nm and the corresponding RS values are $2.64$, $2.44$, $2.17$ and $1.9$, respectively. TS values are noted as $53.83$, $38.06$, $24.81$ and $17.49$ GPa. (see supplementary information\dag) When varying $h_\theta$, the RS increased with very low effect on TS (see circle markers with dotted and solid line in Fig.~\ref{fig:SS}(d)). Whereas, variation of $h_t$ effecting both RS and TS in GNRH. The increase in thickness, increases the separation between stress centers and decreases the TS, which lead to early failure and decrease in RS. This finding implies that changing of $h_t$ strongly influence the both RS and TS of GNRHs. The observations concerning the width effect on the mechanical response are in close agreement with the eralier report based on MD simulations\cite{Chu2014}. } \\
When compared to GNRS, GNRH shows higher stress levels (dotted lines in Fig.~\ref{fig:SS}(d)). From the structural point of view, GNRH differs from GNRS in two factors, one is the smoothness of undulations and the second is the closeness between the undulations. The smoothness of circular arcs in GNRH makes the stress to distribute across all the boundary atoms. The increased number of atoms with higher per atom stress values, increase the total stress in the atomic system. In the case of GNRS, the lower number of atoms with higher per atom stress near the peaks of sine curve makes the total stress lower. From Fig.~\ref{fig:SS}(d), TS values for $2.5-1.6-0^\circ$ GNRH and $9-1.6-2.5$ GNRS are 39.72 and 31.93 GPa, respectively, which represents that the smoothness factor accommodates more number of atoms with high stress levels in horseshoe shape design. The closeness between the undulations increase the deflections in the atomic system, which help to avoid the bond stretching and stress rise. These deflections in GNRS and GNRH systems are measured using the standard deviation of the $z-$coordinates $(\bar{Z})$ \cite{Chen2014}, which is defined as $\bar{Z} = \sqrt{\sum_{i=1}^{N} \left(z_i - z_0\right)^2/N}$, where $z_i$ is the $z-$coordinate of the $i^{\text{th}}$ atom and $z_0$ is the averaged $z-$coordinate over $N$ atoms. Fig.~\ref{fig:SS}(c) plots the computed $\bar{Z}$ with respect to strain for the selected GNRS $(9-1.6-2.5)$ and GNRH (parameter set $h_\theta$ ) systems. $\bar{Z}$ initially increases with strain, which represents that the energy of given mechanical straining used to increase the deflections in both GNRH and GNRS systems. After reaching a maximum deflection, given tensile loading starts stretching the atomic system and decreasing the deflections which decrease $\bar{Z}$. However, the magnitude of $\bar{Z}$ for $0^\circ$ GNRH configuration is high compared to $9-1.6-2.5$ GNRS, which supports that the repulsion interactions in GNRH are heavier compared to GNRS. $\bar{Z}$ increasing with an increase in parameter $h_\theta$. $\bar{Z}$ is highest for $45^\circ$ GNRH. \\
The very strong repulsions exist between the semi-circular rings due to the minimum spacing.	 
The very high deflections and smoothness of circular cross-sections helps to avoid the stress concentrations in GNRH, which helps to enhance the mechanical properties. As a total, very high value of RS is noted for GNRH. With increasing the system size (arc length of GNRH), RS tending to converge to a value of $2$, which is about $17\%$ higher than the graphene kirigami design \cite{Hanakata2018,Hanakata2019}, keeping the stress-levels identical.\\

\subsection{Thermal properties}
\begin{figure*}
	\centering
	\includegraphics[width=1.0\linewidth]{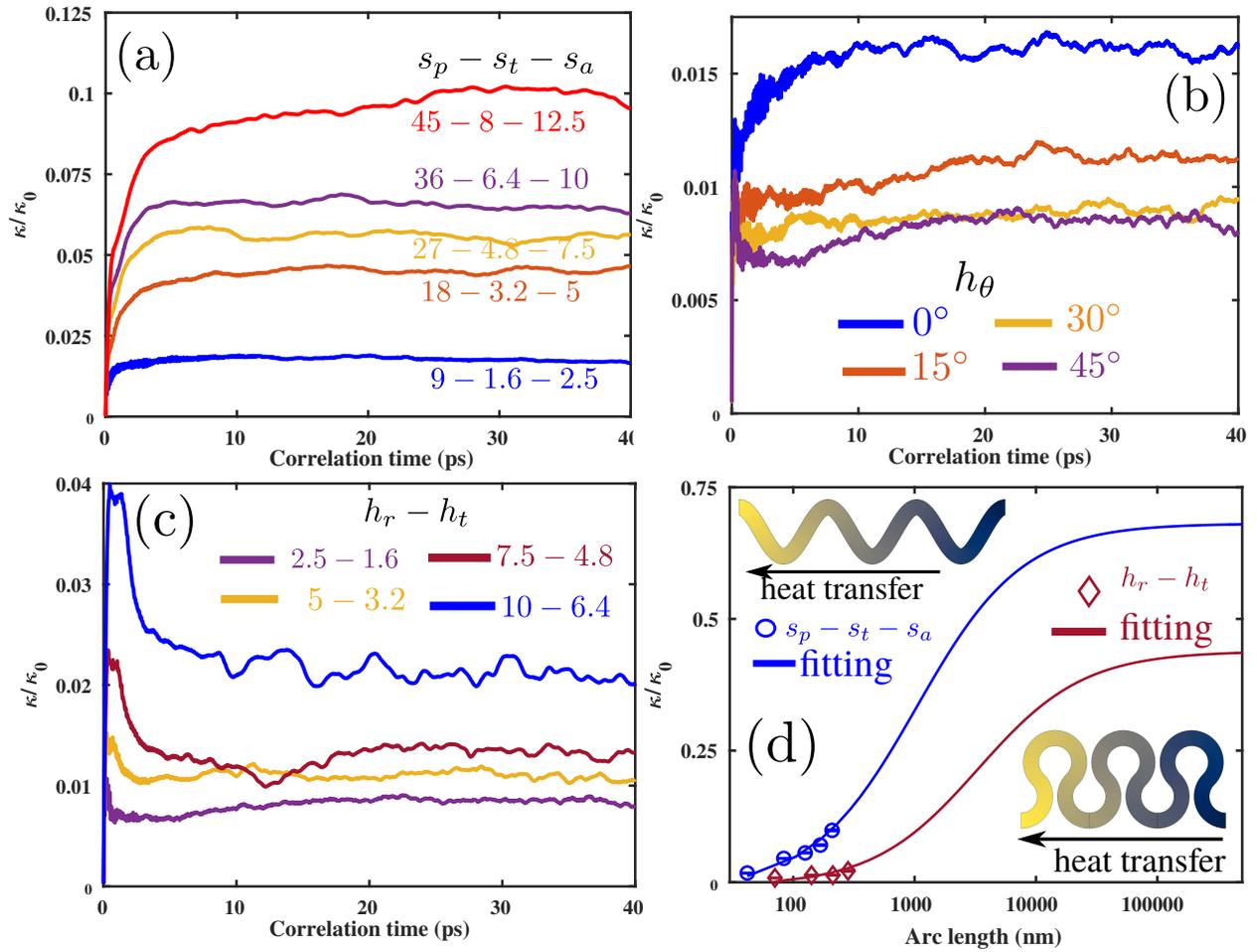}
	\caption{Room temperature thermal conductivity for (a) parameter set $s_p-s_t-s_a$ in GNRS, (b) $h_{\theta}$ and (c) $h_r-h_t$ for GNRH as a function of correlation time. (d) Thermal conductivity for graphene nanosprings at the room temperature as a function of arc lengths. The insets in (d) illustrate the finite element modeling results for temperature distribution of GNRS and GNRH in diffusive regime.}
	\label{fig:TC}
\end{figure*}
We next study the thermal transport through the GNRS and GNRH systems. Fig.~\ref{fig:TC} shows the EMD results of effective thermal conductivity of GNRS for few studied parameter sets, as a function of correlation time. In this case, we normalized the effective conductivity with respect to that of the pristine graphene to better illustrate the suppression rate. As it is clear, for the samples with lower thermal conductivity the convergence occurs at lower correlation times. For $9-1.6-2.5$ GNRS, $\kappa$ is about $0.0175$ times the thermal conductivity of pristine graphene $\kappa_0$ (from Fig.~\ref{fig:TC}(a)). \bj{We employed a square sheet of pristine graphene with $16$ nm side length to estimate $\kappa_0$. The estimated value of $\kappa_0$ is $1000 \pm 100 $ W/(mK), which is an average over $8$ independent simulations. This value is in close agreement with earlier reports using EMD method \cite{Pereira2013}. Note that due to the implementation of inaccurate heat-flux formula for many-body interactions in the LAMMPS tool, EMD method significantly underestimates the the thermal conductivity of graphene \cite{Fan2015}.} The suppressed thermal transport along the GNRS is due to the phonon-boundary scattering in these systems, which is in accordance with the previous reports concerning the graphene kirigami \cite{Mortazavi2017}. \bj{Also, there exist a very strong conversion for the phonon mode polarization from out-of-plane to in-plane and opposite phase for the left and right parts of the unitcell \cite{Yang2012,Zhao2013}. Such conversion of phonon modes loose the transporting energy through the scattering with localized phonon modes near the edges. As a result the transport of heat flux is low and reduce the thermal conductivity.} When $s_t$ increases, the ratio of atoms on the boundaries to the total number of atoms decreases, which results in lower phonon-boundary scattering rate and subsequently facilitate the heat transport.\\	
As proposed in the previous work \cite{Mortazavi2017}, we use a microscale continuum model of graphene spring to evaluate the effective thermal conductivity. This evaluation carried  within the diffusive range, in which the phonon-boundary scattering vanishes. To this aim, a system is modeled within the finite element (FE) approach to establish connections between the effective thermal conductivity and nanoribbon's arc length. We apply inward and outward heat-fluxes on the two opposite sides of GNRS as the boundary conditions. Using the measured temperature gradient along the heat transfer direction, the effective thermal conductivity was computed from the Fourier's law. We then used a first order rational curve fitting to extrapolate the atomistic results (circular markers in Fig.~\ref{fig:TC}(d) that correspond to the averaged $\kappa/\kappa_0$ over several samples of $s_p-s_t-s_a$ and the standard deviation among them) dominated by the phonon-boundary scattering to the diffusive transport by the FE simulations. As shown in Fig.~\ref{fig:TC}(d), this approach could provide a very accurate estimation of thermal transport at different arc lengths, and reveals that the  phonon-boundary scattering starts to vanish at large arc lengths. \\
For GNRH systems with varying $h_{\theta}$ is shown in Fig.~\ref{fig:TC}(b). $\kappa/\kappa_0$ for $2.5-1.6-0^\circ$ GNRH is about $0.0169$, which is nearly same as $9-1.6-2.5$ GNRS. The constant thickness and similar scattering effects in these two spring systems produce the thermal conductivity nearly equal. Keeping thickness constant and increasing the joining angle $h_{\theta}$ to $15^\circ$, $\kappa/\kappa_0$ decreased from $0.0169$ to $0.0111$. As $h_{\theta}$ increases, there is an increase in the radius of curvature of the junction region that connects the two semi-circular segments. The phonon transport through this increased curvature experiences significant scattering, which reduces the heat transfer and $\kappa$. The thermal conductivity for $h_{\theta}$ $30^\circ$ and $45^\circ$ is nearly the same. \\
\bj{We examine thermal conductivity for the GNRH samples used in estimating the effect of width on mechanical properties in Section 3.1. The effective thermal conductivity for $2.5-1.6-45^\circ$ and $5-1.6-45^\circ$ are $0.0084$ and $0.0055$. This represent that increase in $h_r$, increases the radius of curvature and produce more edge localized phonon modes. These modes do not contribute for thermal transport, as a result the thermal conductivity decreases for $5-1.6-45^\circ$ GNRH system (see supplementary information\dag) However, the increase in $h_t$ increase the number of phonon modes in GNRH keeping the density of edge localized modes same. This reduces the edge scattering and increase the phonon transport, thus increase in $\kappa$ \cite{Guo2009}.  }
Fig.~\ref{fig:TC}(c) plots $\kappa$ for increasing values of $h_r$ and $h_t$ keeping joining angle $h_\theta$ as $45^\circ$. As the GNRS radius and thickness increasing, the available region for heat transport increases, which helps to lower the scattering and rise $\kappa/\kappa_0$ from $0.0084$ to $0.0216$. This rise of $\kappa$ is small compared to the GNRS systems due to the large curvature induced scattering. We repeat the FE modeling for GNRH with $h_\theta$ as $45^\circ$ similar to GNRS. The fitting between atomistic results and FE modeling is very good. However, GNRH fitting is converged at significantly larger cut lengths compared to GNRS. This proves that, curvature induced scattering reduces $\kappa$ in GNRH. 
\subsection{Electromechanical properties}	 
\begin{figure*}
	\centering
	\includegraphics[width=1.0\linewidth]{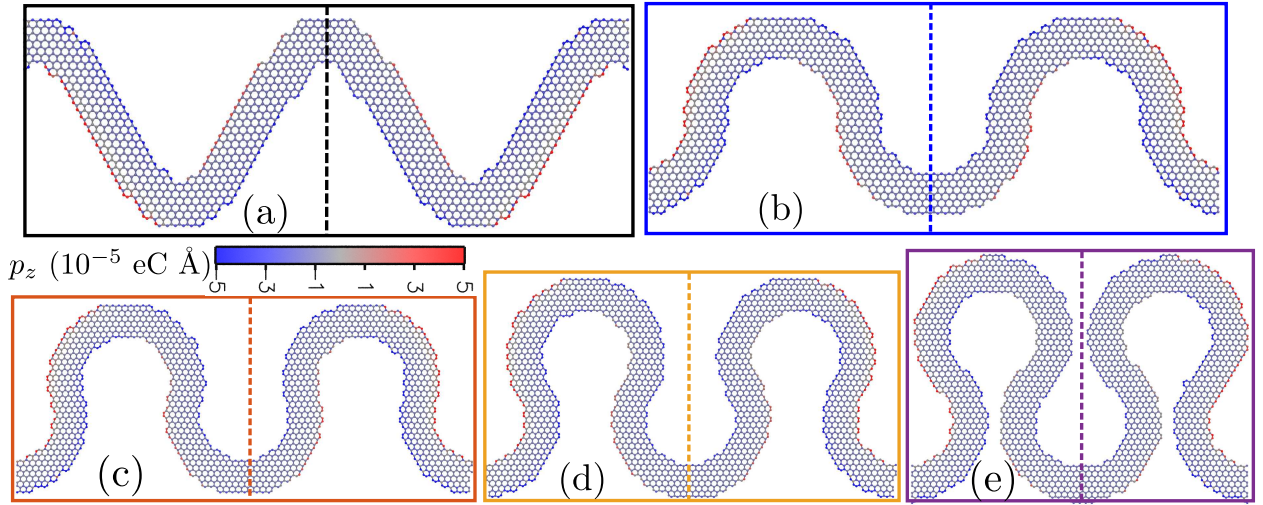}
	\caption{Atomic configurations colored with dipole moment $p_z$ for a strain gradient of $0.006$ nm$^{-1}$. (a) $9-1.6-2.5$ GNRS design. GNRH design with inner radius $2.5$ nm, thickness $1.6$ nm and connecting angle $h_\theta$ as (b) $0^\circ$, (c) $15^\circ$, (d) $30^\circ$ and (e) $45^\circ$. Dashed line indicate the middle portion of the atomic system.}
	\label{fig:pol-ac}
\end{figure*}

The nanoscale electromechanical properties (piezoelectricity and flexoelectricity) have been gaining intense attention due to their ability to sustain large deformations. This feature adds many different applications in the energy conversion process. These properties are limited in pristine graphene due to the crystallographic symmetry. The structural and chemical modifications break the symmetry and induce polarization under mechanical deformations \cite{Ong2012,Chandratre2012,Xue2012,Ahmadpoor2015,Kundalwal2017,Javvaji2018,Ghasemian2020}. The bending deformation of pristine graphene induces a polarization due to the change of hybridization state of the carbon atom, known as pyramidalization \cite{Dumitrica2002,Ahmadpoor2015,Zhuang2019}. In GNRS and GNRH systems, cut patterns cancel the symmetry and promises for electromechanical coupling. In this section, we subject the GNRS and GNRH system to both tensile and bending deformations to obtain the respective polarization variations. \\
To calculate the polarization in the atomic system, we utilize the charge-dipole (CD) model along with the short-range bonded interactions (Tersoff potential). According to CD model, each atom $i$ is associated with charge $q_i$ and dipole moment $\mathbf{p}_i$ \cite{Mayer2005,Mayer2007}. The mathematical CD formulation involves the various contributions from charge-charge, charge-dipole and dipole-dipole interactions to the total system short-range interaction energy. The minimization of energy function gives the numerical values of $q_i$ and $\mathbf{p}_i$. The complete details about the CD model and estimation of charges and dipole moments can be found in \cite{Javvaji2018,Zhuang2019} and references therein. \\    
For applying deformation, we have added left and right rectangular regions to the GNRS and GNRH systems by discarding the periodic boundary condition used in Section 2. These regions have equal $s_t$ or $h_t$ to the spring systems with $1$ nm length along the spring longitudinal direction. The left and right regions helps to hold the given displacement, particularly during bending test, and relax the remaining system. We define a load cycle by prescribing the displacement of atoms to left and right regions for $1$ ps time period, followed by a relaxation of $2$ ps time period. Because of the non periodic boundaries, we perform simulations at different repetitions of sinus and horseshoe shapes in spring systems. These simulations help us to study the size effect on electromechanical properties. For every load cycle, we note the evolution of atomic configuration, corresponding charge and dipole moments. The total polarization of the atomic system is the sum of all atomic dipole moments divided by the volume of atomic system. \\
\begin{figure*}[h]
	\centering
	\includegraphics[width=1.0\linewidth]{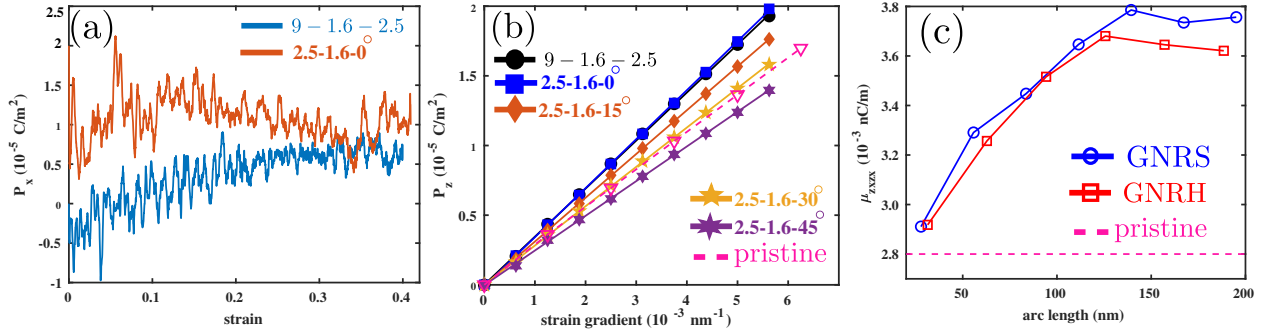}
	\caption{(a) Variation of polarization $P_x$ with tensile strain. (b) Bending induced polarization $P_z$ response with respect to strain gradient. (c) Dependence of flexoelectric coefficient $\mu_{zxzx}$ with arc length of GNRS and GNRH systems. Legends indicate the respective atomic configurations used.}
	\label{fig:pol}
\end{figure*}
For tensile deformation, we apply a displacement in longitudinal direction $u_x=\dot{\epsilon} t_{\text{load}} l_0$ to the atomic system, where $\dot{\epsilon}$ is strain rate  equal to  1$\times$10$^8$ s$^{-1}$ as used in Section 2, $t_{\text{load}}$ is the loading time (1 ps) and $l_0$ is the initial length of the atomic system in longitudinal direction. The load cycles continued to reach the strain $\epsilon_{xx}$ of $0.4$ for $9-1.6-2.5$ and $2.5-1.6-0^\circ$ systems. The strain limit $0.4$ corresponds to the linear rise in stress-strain response for these systems as shown in Fig.~\ref{fig:SS}(a). At each load cycle, the polarization is measured and the variation with strain is plotted in Fig.~\ref{fig:pol}(a). The variation in polarization with strain is nearly negligible for both GNRS and GNRH systems. The coefficient of variation for the polarization response is nearly equal to 1 for both GNRS and GNRH, similar to non-piezoelectric pristine graphene \cite{Javvaji2018}. The cancellation of polarization at the sinus and horseshoe cut patterns make these systems as non-piezoelectric materials. \\
For bending deformation, we supply the following out-of-plane displacement field to the atomic system 
\begin{equation}
u_z = K \frac{x^2}{2},
\label{eq:uz}
\end{equation}
where $x$ represent the atom coordinate in the longitudinal direction, $K$ represent the inverse of curvature (strain gradient) of the bending plane. After prescribing the bending deformation, the atoms belongs to the left and right termination regions  are held fixed. Whereas, the interior atoms are allowed to relax to energy minimizing positions using the conjugate-gradient algorithm. For the energy minimized configuration, we note the charges and dipole moments for each atom. From these, the relationship between polarization to strain and strain gradient as
\begin{equation}
P_z = d_{zxz} \epsilon_{xz} + \mu_{zxzx} \frac{\partial \epsilon_{xz}}{\partial x},
\label{eq:pzt}
\end{equation}
where $P_z$ is the out-of-plane polarization, $\mu_{zxzx}$ is the out-of-plane bending flexoelectric coefficient, $d_{zxz}$ is the piezoelectric coefficient and $\epsilon_{xz}$ is the strain component. Here the piezoelectric contribution $(d_{zxz}\epsilon_{xz})$ is removed from the total polarization, because the given bending deformation leads to a linear variation in $\epsilon_{xz}$ along the $x$ direction. The linear variation demonstrates that the induced deformation is symmetric and the resulting polarization due to strain is canceled out \cite{Zhuang2019}. From Eq.~\ref{eq:uz}, the strain gradient is given as
\begin{equation}
\frac{\partial \epsilon_{xz}}{\partial x} = \frac{1}{2} \frac{\partial}{\partial x} \left( \frac{\partial u_z}{\partial x} + \frac{\partial u_x}{\partial z} \right) = \frac{1}{2} \frac{\partial^2 u_z}{\partial x^2} = \frac{1}{2} K = K_{\text{eff}}, 
\label{eq:sg}
\end{equation}
where $u_x$ is zero because of fixing the atom positions belongs to the left and right boundary and $K_{\text{eff}}$ is the effective strain gradient, which is equal to half of the given value of $K$. Substituting Eq.~\ref{eq:sg} along with zero piezo contribution in Eq.~\ref{eq:pzt} leads to
\begin{equation}
P_z = \mu_{zxzx} K_{\text{eff}}.
\label{eq:pz}
\end{equation}
The polarization $P_z$ at various strain gradients $K_{\text{eff}}$ is plotted in Fig.~\ref{fig:pol}(b). The dipole moment $\mathbf{p}_i$ on atom rises due to the  pyramidalization. The bending deformation transforms the atomic hybridization of carbon atom from $sp^2$ to $sp^3$. The bending of the bond between carbon atoms forces the valence electron to develop a bonding interaction with neighboring atoms. This interaction allows a mixing between valence $(\pi)$ and bonding $(\sigma)$ electrons which lead to $\pi-\sigma$ interactions. The $\pi-\sigma$ interaction modifies the charge state of carbon atoms and the local electric fields, which leads to the dipole moment of that atom. The increased deformation increases the $\pi-\sigma$ interaction, which rise the dipole moment. Fig.~\ref{fig:pol-ac}(a) and (b) show the distribution of dipole moment for the unitcells next to the center of the atomic systems. Both sides of the lateral edges have opposite dipole moments, and their magnitude is a function of bending  displacement (from Eq.~\ref{eq:uz}). The observed linear variation between polarization and strain gradient for spring systems is similar to the pristine graphene variation. The slope of this variation gives the flexoelectric coefficient $\mu_{zxzx}$. Numerical value of $\mu_{zxzx}$ for $9-1.6-2.5$ and $2.5-1.6-0^\circ$ systems are $0.0034$ and $0.0035$ nC/m, which is $25\%$ higher than pristine graphene \cite{Zhuang2019}. The hybridization angle $\theta_{\sigma\pi}$ is the angle between a fixed  out-of-plane point to one of the bond between carbon atoms. For the initial or flat system, this angle is exactly $90^\circ$. The deformation of bond changes this angle. The change in angle for pristine and $2.5-1.6-0^\circ$ system are $4.62$ and $6.38^\circ$, respectively. The increment in $\theta_{\sigma\pi}$, increases the dipole moment of the system \cite{Dumitrica2002}, which increase the flexoelectric coefficient. Note that we calculated the hybridization angle at the same location in both the systems. \\
From Fig.~\ref{fig:pol}(b), the slope of the polarization to strain gradient curve is decreasing with the increase of $h_{\theta}$ in GNRH systems. The ratio of change in pyramidalization angle in GNRH to pristine grpahene decreases as $1.27$ , $1.14$ and $1.05$ and $0.85$ with $h_\theta$ as $0^\circ$, $15^\circ$, $30^\circ$ and $45^\circ$, respectively. The decrease in $\theta_{\sigma\pi}$ decreases, the dipole moment distribution, as seen from Figs.~\ref{fig:pol-ac}(c) to (d), which decreases the flexoelectric coefficient.  For $h_{\theta}=45^\circ$, atoms on the lateral boundaries near the central line does not have dipole moments. The strong repulsions between these edges cancels the effect of pyramidalization, which decreases the flexoelectric coefficient. \\
\bj{In order to check the dependence of $h_t$ on the polarization, we consider $2.5-2.4-0^\circ$ and $2.5-3.2-0^\circ$ GNRH configurations. For these configurations, the polarization variation and flexoelectirc coefficient are nearly equal to the $2.5-1.6-0^\circ$ GNRH. The increase in thickness does not effecting the induced polarization and flexoelectiric coefficient (see supplementary information\dag). Since the increased thickness unable to change the pyramidalization angle further, makes the polarization comparable with smaller thickness system.}\\
Further, the result of flexoelectric coefficient with length variation is given in Fig.~\ref{fig:pol}(c) for GNRS and GNRH $(h_{\theta}=0^\circ)$. This result represent that there is a boundary effect when arc length is less than $100$ nm. For systems with arc length about $30$ nm, the local electric fields is strongly effected by the left and right region of atoms. Where the imposed boundary condition constrain the natural motion of the interior atoms, which restricts the process of pyramidalization and controls the dipole moment. When increasing the arc length this effect is slowly nullifying and the atomic configuration deforms to generate the dipole moments. For systems with lengths higher than $100$ nm, the boundary effect if completely negligible and the flexoelectric coefficients turn into a stable value. Finally, the flexoelectric coefficient of GNRS and GNRH-$0^\circ$ system is $0.25$ times higher than the pristine graphene. 
\section{Conclusion}
Motivated by a latest experimental advance, we performed extensive classical molecular dynamics simulations to explore the mechanical, thermal conductivity and electromechanical properties of graphene nanoribbon springs (GNRS). In particular, we examined the effects of different GNRS design parameters on their physical properties. We found that by optimal design of GNRS systems, they can yield higher stretchability in comparison with kirigami counterparts. Horseshoe shape design of GNRS were found to show the better stretchability in comparison with other design strategies. In the aforementioned case, large deflections due to the strong repulsions between semi-circular rings help to keep the load bearing at larger strain levels. Our analysis of deformation process reveals that the stress concentrations occurring near the peak portions of GNRS induce local failure of carbon bonds and lead to final failure of structure. The thermal conductivity of GNRS were found to be substantially dependent on the width of nanoribbon's width, due to the fact that phonon boundary scattering dominates the thermal transport. On this basis, we could establish the connection between the effective thermal conductivity of GNRS as a function of nanoribbon's width size, by extrapolating the molecular dynamics results to the diffusive heat transfer model by the finite element. This approach can be used to effectively estimate the thermal conductivity, but also suggest that  the thermal transport of GNRS can be effectively tuned by changing the design parameters. The negligible variation between polarization and strain proves that GNRH and GNRS systems are non-piezoelectric similar to graphene. A linear variation of polarization with strain gradient is observed in bending deformation test. The flexoelectric coefficient for GNRS and GNRH-$0^\circ$ is $25\%$ higher than graphene. The decrease in $\mu_{zxzx}$ with increasing $h_{\theta}$ is due to the decrease of pyramidalization angle. Our extensive theoretical results highlight the superior stretchability, finely tunable thermal transport and improved flexoelectric coefficient of GNRS, and suggest them as highly attractive components to design flexible nanodevices. The obtained results will hopefully guide future theoretical and experimental studies, to extend the idea of nanoribbon springs for graphene and other 2D materials. 

\section*{Conflicts of interest}
There are no conflicts to declare.

\section*{Acknowledgments}
The authors gratefully acknowledge the sponsorship from the ERC Starting Grant COTOFLEXI (No. 802205). We acknowledge the support of the cluster system team at the Leibniz Universit\"{a}t of Hannover, Germany in the production of this work. 




\bibliography{references} 
\bibliographystyle{ieeetr} 

\end{document}